\documentclass[preprint,amssymb]{revtex4-1}
\usepackage{graphicx} 
\usepackage{dcolumn}  
\usepackage{bm}       
\usepackage{wasysym}  
\usepackage{courier}  
\usepackage{braket}   
\usepackage{color}    

\begin{document}

\title{Theory of atomic scale quantum dots in silicon: dangling bond quantum dots on silicon 
surface}

\author{Alain Delgado\textsuperscript{1}, Marek Korkusinski\textsuperscript{1,2},
Pawel Hawrylak\textsuperscript{1}}

\affiliation{\textsuperscript{1}Department of Physics, University of Ottawa, Ottawa, Canada K1N
6N5}

\affiliation{ \textsuperscript{2}National Research Council of Canada, Ottawa, Canada K1A OR6 }

\begin{abstract}

We present a theory and a computational tool, Silicon-{\sc Qnano},  describing atomic 
scale quantum dots in silicon. The developed methodology is applied to model dangling bond quantum dots 
(DBQDs) created on a passivated H:Si-(100)-(2$\times$1) surface by removing a hydrogen atom. 
The electronic properties of the DBQD are computed by embedding it in a computational box of 
silicon atoms. The surfaces of the computational box were constructed by using density functional theory as 
implemented in the {\sc Abinit} package. The top layer was reconstructed by the formation of Si 
dimers passivated with H atoms while the bottom layer remained unreconstructed and fully saturated 
with H atoms. The computational box Hamiltonian was approximated by a tight-binding (TB)
Hamiltonian by expanding the electron wave functions as linear combinations of atomic orbitals and 
fitting the bandstructure to {\it ab-initio} results. The parametrized TB Hamiltonian was used to 
model large finite Si(100) boxes (slabs) with number of atoms exceeding present capabilities of {\it 
ab-initio} calculations. The removal of one hydrogen atom from the reconstructed surface resulted 
in a DBQD state with a wave function strongly localized around 
the Si atom and the energy in the silicon bandgap. The DBQD could be charged with zero, one, and two 
electrons. The Coulomb matrix elements were calculated and the charging energy of a two electron 
complex in a DBQD obtained.
\end{abstract}

\maketitle

\section{INTRODUCTION}
\label{sec_intro}
There is currently interest in extending silicon-based microelectronics to quantum technologies, 
including silicon nanocrystals \cite{priolo-gregorkiewicz-nature-2014}, gated quantum dots 
\cite{watson-vandersypen-nature-2018,studenikin-sachrajda-apl-2018}, and 
dopants  \cite{dassarma-koiller-sscom2004,salfi-simmons-nature-2014}. 
Recently, several groups have demonstrated the possibility of using scanning tunneling microscopy (STM) to 
remove hydrogen atoms from a hydrogen passivated Si(100)-(2x1) surface 
\cite{wolkow-bruno-2014-nanocomputing,haider-wolkow-prl-2009,taucer-wolkow-prl-2014,livadaru-wolkow-njp-2010,doumergue-baratoff-prb-1999,schofield-bowler-nature-2013,lorente-nanotechnology-2014, wyrick-silver-nl-2018}.
The removal of a hydrogen atom 
from the surface creates a dangling bond (DB) in a silicon atom with corresponding energy in the gap 
of bulk Si, well below the bottom of the conduction band \cite{haider-wolkow-prl-2009,raza-raza-prb-2007}. 
This DB can be charged in a controlled way with electrons drawn from n-type doped Si substrate \cite{wolkow-bruno-2014-nanocomputing, taucer-wolkow-prl-2014}. 
DBs were used for atom-by-atom construction of linear chains and cyclic artificial molecules in silicon
 \cite{taucer-wolkow-prl-2014, doumergue-baratoff-prb-1999, schofield-bowler-nature-2013,lorente-nanotechnology-2014, wyrick-silver-nl-2018}.

When the dangling bond quantum dot (DBQD) is charged and/or manipulated, the quantum structure involves a large number of 
atoms. The same is true for quantum circuits created with dopants, nanocrystals, and gated silicon 
quantum dots - the number of silicon atoms involved even in a very small circuit can easily exceed 
a million. Hence to develop an understanding of atomic scale  quantum 
computing devices  in silicon one needs a computational tool suitable for 
designing circuits made of millions of atoms. 
Here we describe Silicon-{\sc Qnano} (Si-{\sc Qnano}), a {\sc Qnano} computational platform 
\cite{sheng-hawrylak-prb-2005, zielinski-hawrylak-prb-2010, korkusinski-voznyy-prb-2011} for the 
design of quantum nanostructures in silicon and apply it to atomic scale DB-based quantum 
dots  on a surface of silicon.

\section{Computational silicon box with reconstructed  surface}
\label{surface_reconstruction}
\begin{figure}[t]
\begin{center}
\includegraphics[scale=1,angle=0]{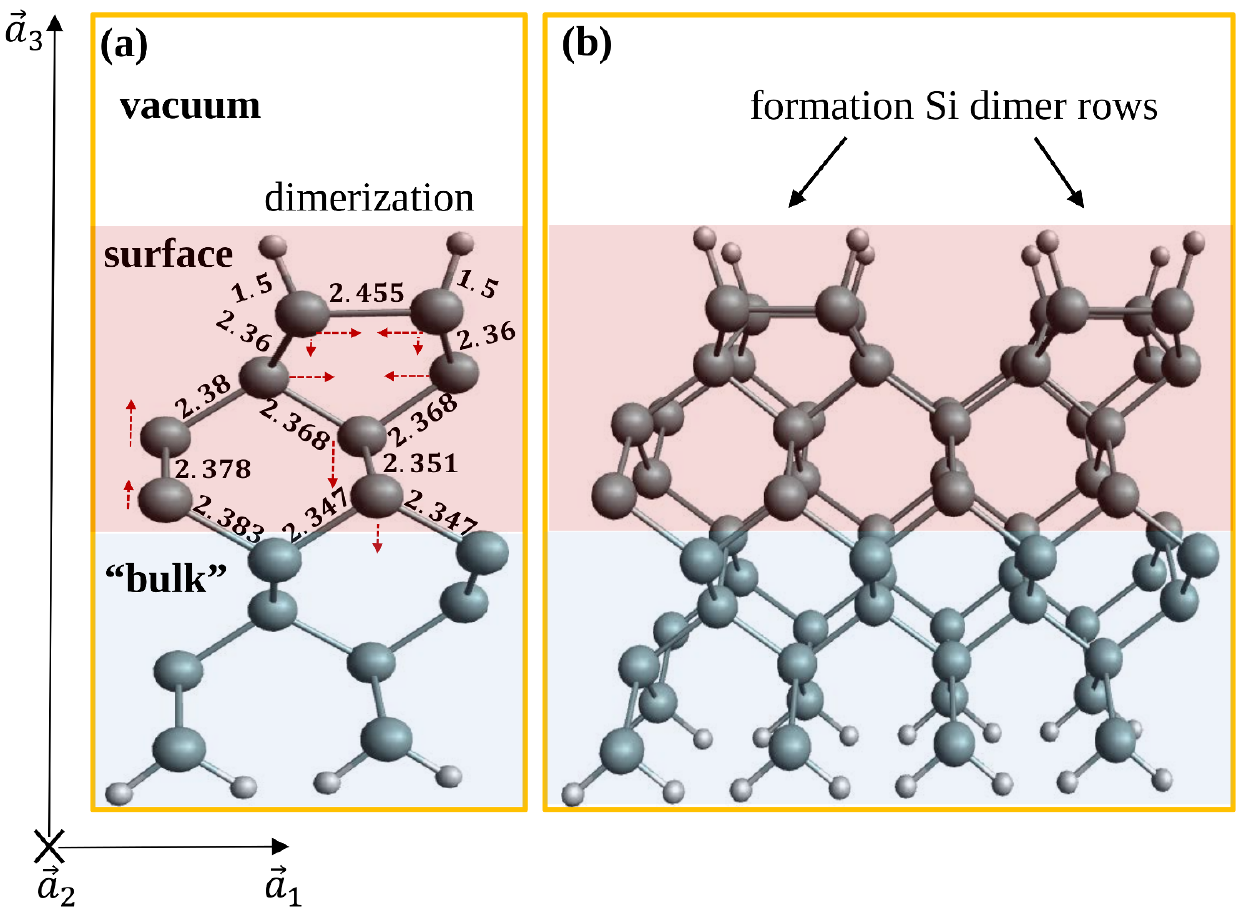}
\caption{ \label{fig_opt_scell} (color online) (a) Optimized atomic structure of the supercell used 
to model the top surface reconstruction of a Si-(100) slab. Bond lengths are reported explicitly in 
Angstroms. Dashed (red) arrows indicate the displacements of Si atoms respective to their positions 
in bulk material due to surface reconstruction. (b) A larger slab supercell illustrating the 
formation of rows of Si dimers on the reconstructed surface of the slab.}
\end{center}
\end{figure}

In this section we describe a finite  computational box made of silicon atoms with a reconstructed and 
passivated  top  surface on which impurities, defects, dangling bonds, or external gates are 
implemented. The bottom surface, on the other hand, is constructed to  simulate bulk silicon. 
We start with a small box with a number of silicon atoms suitable for {\it ab-initio} calculations.
Fig. \ref{fig_opt_scell}(a) shows the computational supercell used to model such a box, or a slab, here with
$8$ Si layers. The top Si surface, passivated with H atoms,  aims at simulating a real Si surface, followed by a vacuum region in the [0,0,1] direction.  This supercell is repeated periodically in the lateral and vertical directions using the 
lattice vectors ${\vec a}_1=a_1 {\hat {\bf x}}$, ${\vec a}_2=a_2 {\hat {\bf y} }$ and 
${\vec a}_3=a_3 {\hat {\bf z} }$. Here $a_1=\sqrt{2}a_\mathrm{Si}$, $a_2=\sqrt{2}a_\mathrm{Si}/2$ 
and $a_3=2a_\mathrm{Si} + h_\mathrm{vac}$ where $a_\mathrm{Si}$ is the lattice constant 
of bulk Si and $h_\mathrm{vac}$ is the height of the vacuum region. 

In order to understand the structure and electronic properties of our Si computational box we perform density functional theory (DFT) calculations using the PBE generalized gradient 
approximation to the exchange-correlation energy functional \cite{pbe_functional_original, 
pbe_functional_erratum} as implemented in the {\sc Abinit} code \cite{Gonze-zwanziger-cpc-2009}. 
 All calculations were 
performed using a plane wave basis set,  truncated with a kinetic energy cutoff of $20$ Ha ($544$ eV). 
A grid with 6 $\times$ 12 $\times$ 1 k-points was used for Brillouin zone integrations using 
Monkhorst-Pack method \cite{monkhorst-pack-prb-1976}. 
First, we optimized the lattice constant of bulk silicon and obtained $a_\mathrm{Si}=5.46$ 
\AA. Note that this value  agrees up to $0.03$ \AA{} with the experimentally observed  lattice 
constant of Silicon, $a_\mathrm{Si}^\mathrm{exp}=5.43$ \AA{} \cite{madelung-otfried-semiconductors}. 

Next we proceed to the surface reconstruction. 
First, 
for a given number of Si layers the thickness $h_\mathrm{vac}$ of vacuum above the surface was varied until 
total energy did not depend on it.  We found that a vacuum region of $16$ \AA{} was enough to suppress 
interactions between periodic images in the ${\hat {\bf z} }$ direction.
Next, 
we varied the number of Si layers of the slab 
and found that  $8$ layers were sufficient to achieve convergence of the surface energy per 
surface unit cell. With $8$ Si layers there are $16$ Si atoms and $6$ H atoms in a supercell,  as shown in 
 Fig. \ref{fig_opt_scell}(a). 

Surface reconstruction was achieved by minimizing the total energy with respect to the 
atomic positions of H and Si atoms in the top four layers of the slab. Atomic coordinates were 
adjusted until the maximum interatomic force was less than $0.001$ eV$/$\AA. 
In Fig. \ref{fig_opt_scell}(a) we illustrate with arrows the displacements of Si atoms with respect to their 
positions in the bulk material. 
The positions of Si atoms in the bottom four layers were not optimized 
as relaxation decreases very rapidly as we move away from the top 
surface. The bottom surface is unreconstructed and fully passivated
with H atoms to simulate a seamless transition to the bulk material.

The essence of the reconstruction can be understood by comparing the
positions and coordination of the two topmost surface Si atoms in
the unit cell in  Fig.~\ref{fig_opt_scell}(a) to the two Si atoms at
the bottom of the cell, simulating bulk material.
Each of the two bottom Si atoms is bonded to two other atoms above it,
of which only one is seen in Fig.~\ref{fig_opt_scell}(a).
The bonding becomes more apparent when we begin building up the slab,
as presented in Fig.~\ref{fig_opt_scell}(b).
Moreover, each Si atom of the bottom layer is connected to two H atoms
underneath it.
In the full slab geometry, the bottom layer Si atoms are distributed
regularly and form the (100) crystal plane as they do in the bulk. 
In contrast, the surface (top) Si atoms are coordinated differently.
Each atom has {\em three} Si nearest neighbors, and a single DB, which is saturated by the H atom.
As indicated by the red arrows in Fig.~\ref{fig_opt_scell}(a),
this reconstruction is realized by 
(i) a shift of atomic positions of the first (top) and, to a lesser extent, the second layer, and
(ii) the formation of a new Si-Si bond between the top two Si atoms.
We stress that the new bond is remarkably only about $5$\% longer
than the bulk Si-Si bond.
The key qualitative change introduced by the surface reconstruction is
that this bond connects only alternate Si pairs, resulting in dimers
evident in Fig.~\ref{fig_opt_scell}(b).

\section{Tight-binding electronic-structure calculations}
\label{sec_model}
In this section we describe the model tight binding (TB) Hamiltonian we use to perform approximate calculations of the 
Kohn-Sham (KS) quasiparticles of the Si slab described in the previous section.  The KS quasiparticle Hamiltonian
reads:
\begin{equation}
  \hat{H}_{QP} = \frac{{\vec p}^{~2}}{2m} 
  + V_\mathrm{atoms}(\vec{r})+ V_\mathrm{Hartree}(\vec{r}) + V_\mathrm{xc}(\vec{r}),
\label{eq_qp_hamilt}      
\end{equation}
where $V_\mathrm{atoms}(\vec{r})$ is the sum of atomic potentials, $V_\mathrm{Hartree}(\vec{r})$ is 
the Hartree potential produced by all electrons, and $V_\mathrm{xc}(\vec{r})$ is the 
exchange-correlation potential. If we  carry out the fully self-consistent DFT calculations as in the previous section, the KS Hamiltonian is expressed in terms of 
atomic, Hartree, and exchange-correlation potentials, themselves functionals of the 
ground state electronic density. Since we do not know the Hamiltonian for the number of atoms exceeding {\em ab initio} capabilities, we parametrize it  
in a TB form by expanding the electron wave function as a linear combination of 
atomic orbitals (LCAO) of the type $\alpha$ on the atom at position ${\vec R}$:
\begin{equation}
\vert \phi \rangle=\sum_{\vec{R},\alpha}c_{\vec{R}\alpha}|\vec{R}\alpha\rangle.
\label{eq_lcao}
\end{equation}
In our tight-binding approach we retain ten valence orbitals 
for each Si atom: one $s$, three  $p$, five  $d$,  and one additional $s^*$ 
orbital that accounts for higher lying states,  and  by one $s$ orbital on each H atom. 
In this basis, the TB Hamiltonian can be written in second quantization 
as follows:
\begin{equation}
  \hat{H}_\mathrm{TB} =
  \sum_{i=1}^{N_\mathrm{Si} + N_\mathrm{H}} \sum_{\alpha=1}^{N_\mathrm{orb}^{(i)}}
       \varepsilon_{i\alpha}c_{i\alpha}^+c_{i\alpha}
+   \sum_{i=1}^{ N_\mathrm{Si}+ N_\mathrm{H} } ~ \sum_{j=1}^{\mathrm{NN}^{(i)}} 
~\sum_{\alpha=1}^{N_\mathrm{orb}^{(i)}} ~ \sum_{\beta=1}^{N_\mathrm{orb}^{(j)}} 
       t_{i\alpha,j\beta}c_{i\alpha}^+c_{j\beta}.
\label{eq_tb_hamilt}
\end{equation}
In Eq. (\ref{eq_tb_hamilt}), $c_{i\alpha}^+$ ($c_{i\alpha}$) is the creation (annihilation) operator 
of an electron on the orbital $\alpha$ localized on the site $i$, $\varepsilon_{i\alpha}$ is the 
corresponding on-site energy, and $t_{i\alpha,j\beta}$ describe the hopping of the particle between 
orbitals on neighboring sites. $N_\mathrm{Si}$ and $N_\mathrm{H}$ are, respectively, the total 
number of Si and H atoms in the slab, $N_\mathrm{orb}^{(i)}$ is the number of atomic orbitals 
centered on site $i$ and $\mathrm{NN}^{(i)}$ is the number of nearest neighbors of the $i$-th atom, 
that is, $4$ for Si atoms and $1$ for H atoms.

The off-diagonal matrix elements (hopping parameters) of our Hamiltonian are calculated according to 
the Slater-Koster rules \cite{slater-koster}.
In this approach, the hopping parameters
$t_{i\alpha,j\beta}$ are expressed as geometric functions of two-center integrals and depend only on 
the relative positions of the two centers $i$ and $j$. Contributions from three-center integrals are 
neglected. A detailed explanation of how to evaluate tunneling matrix elements was 
already published in Refs. [\onlinecite{zielinski-hawrylak-prb-2010},\onlinecite{slater-koster}]. 
Here the on-site energies $\varepsilon_{\alpha}$ and tunneling matrix elements $t_{i\alpha,j\beta}$ 
are not directly calculated, but obtained by fitting the TB band structure to the respective values measured experimentally or obtained by first-principles calculations. In this work, we use 
our own sets of TB parameters that fit the {\it ab-initio} DFT band structure of the passivated Si 
slab with reconstructed surface. More details about the optimized set of TB parameters are given
in Sections \ref{subsec_bandsHsurface} and \ref{subsec_dbwires}.

\section{Band structure of a S\lowercase{i} box with Hydrogen passivated and reconstructed surface 
- a tight binding model }
\label{subsec_bandsHsurface}

\begin{figure}[h]
\begin{center}
\includegraphics[scale=1.2,angle=0]{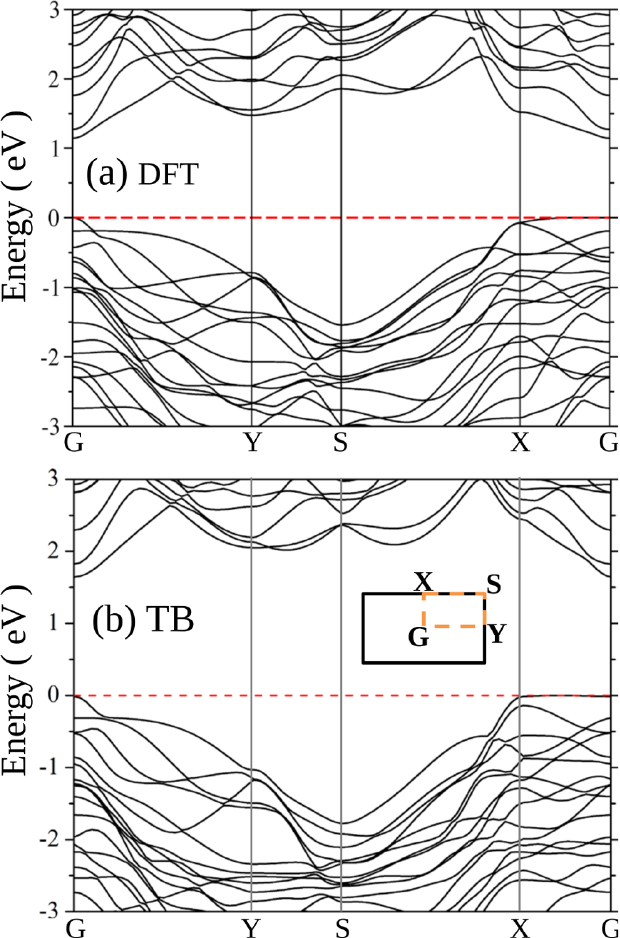}
\caption{ \label{fig_bands_passsurface} (color online) Band structure of the reconstructed and 
H-passivated Si slab shown in Fig. \ref{fig_opt_scell} calculated with (a) DFT using 
GGA-PBE approximation and plane waves as implemented in the {\sc Abinit} program, and (b) the
TB method with the set of parameters (see Table \ref{table_bulk_tb_parameters}) optimized 
to reproduce the {\it ab-initio} DFT band structure. The Fermi level
is indicated with the dashed red line.}
\end{center}
\end{figure}

In this section we report results and comparison of TB 
and DFT calculations of the band structure of the 
Si box (slab) shown in Fig. \ref{fig_opt_scell}. 
In Fig. \ref{fig_bands_passsurface}(a) we plot the energy bands of the model box calculated with {\sc Abinit} along the path defined by the symmetry points $\mathrm{G}=(0,0,0)$, $\mathrm{Y}=(0, \pi/a_2, 0)$, $\mathrm{S}=(\pi/a_1, \pi/a_2, 0)$ and 
$\mathrm{X}=(\pi/a_1, 0, 0)$ of the surface Brilloin zone. The supercell shown in Fig. 
\ref{fig_opt_scell}(a) has 16 Si atoms and 6 H atoms. With pseudopotentials  
accounting for core electrons, we have a total number of $N_e=70$  valence electrons that occupy the 
first 35 spin-degenerate lowest-energy bands. The Fermi level in Fig. \ref{fig_bands_passsurface} 
is indicated by the horizontal dashed red line. The energy gaps at the G, Y, S and X points obtained 
with DFT calculations are reported in Table
\ref{table_gaps_passivated}. 
\begin{table}[b]
\caption{Energy gaps between the top of the valence band and the bottom of the 
conduction band ($E_\mathrm{gap}$) plotted in Fig. \ref{fig_bands_passsurface} calculated with DFT 
and TB methods at the points G, Y, S and X of the surface Brilloin zone. All values are 
given in eV.}
\label{table_gaps_passivated}
\begin{tabular}{ll|l}
\hline\noalign{\smallskip} 
  & $~~~~~$ DFT & $~~~~~~$ TB \\
\noalign{\smallskip}\hline\noalign{\smallskip}
$E_\mathrm{gap}(\mathrm{G})$ & $~~~~$ 1.144 $~~~~$   & $~~~~$ 1.656 $~~~~$ \\
                             &                       &                     \\
$E_\mathrm{gap}(\mathrm{Y})$ & $~~~~$ 2.262 $~~~~$   & $~~~~$ 3.078 $~~~~$ \\
                             &                       &                     \\
$E_\mathrm{gap}(\mathrm{S})$ & $~~~~$ 3.401 $~~~~$   & $~~~~$ 4.139 $~~~~$ \\
                             &                       &                     \\
$E_\mathrm{gap}(\mathrm{X})$ & $~~~~$ 1.592 $~~~~$   & $~~~~$ 2.496 $~~~~$ \\
\noalign{\smallskip}\hline
\end{tabular}
\end{table}
We  note that the DFT band structure in Fig. \ref{fig_bands_passsurface}(a) reproduces very 
well the results reported recently by Bohloul {\it et al.} \cite{bohlou-guo-nanoletters-2016} using LDA and the projector augmented wave method 
implemented in {\sc Vasp}. \cite{kresse-furthmuller-prb-1996} We  also note that DFT 
calculations performed within the LDA approximation by Wang {\em et al.} \cite{wang-pollmann-prb-2006} predict a larger electronic gap of $2.0$ eV at the G point 
for a Si slab with a smaller lattice constant.
The lower panel of Fig. \ref{fig_bands_passsurface} shows the energy bands for the same slab 
obtained by diagonalizing the TB Hamiltonian in Eq. (\ref{eq_tb_hamilt}) assuming 
periodic boundary conditions. 

\begin{table}[h]
\caption{Tight binding parameters for Si and H atoms on reconstructed surface used in the TB Hamiltonian to calculate the band structure shown in Fig. \ref{fig_bands_passsurface}(b). All values are in eV.}
\label{table_bulk_tb_parameters}
\begin{tabular}{lll|lll}
\hline\noalign{\smallskip}
\multicolumn{3}{c}{$~~~~~$onsite energies} & \multicolumn{3}{c}{$~~~~$hopping parameters} \\ 
\hline\noalign{\smallskip} 
  & Si & H &  & Si-Si & Si-H \\
\noalign{\smallskip}\hline\noalign{\smallskip}
$\varepsilon_s~$    & -2.152 $~~~~$   & -1.0  & $~t_{ss}~~~~~$     & -1.959 $~~~~$  & -1.959 \\
$\varepsilon_{p_{x,y}}~$  & 4.229          & -     & $~t_{sp}$          &  3.026   & -       \\
$\varepsilon_{p_z~}$      & 4.229   & -     & $~t_{ps}$          &  3.026   & 3.026   \\
$\varepsilon_{d~}$        & 13.789  & -     & $~t_{pp\sigma}$    &  4.104   & -       \\
$\varepsilon_{s^*~}$      & 19.117  & -     & $~t_{pp\pi}$       & -1.518   & -       \\
                          &         &       & $~t_{sd}$          & -2.285   & -       \\
                          &         &       & $~t_{pd\sigma}$    & -1.355   & -       \\
                          &         &       & $~t_{pd\pi}$       &  2.385   & -       \\
                          &         &       & $~t_{dd\sigma}$    & -1.681   & -       \\
                          &         &       & $~t_{dd\pi}$       &  2.588   & -       \\
                          &         &       & $~t_{dd\delta}$    & -1.814   & -       \\
                          &         &       & $~t_{s^*s}$        & -1.522   & -       \\
                          &         &       & $~t_{s^*p}$        &  3.156   & -       \\
                          &         &       & $~t_{s^*d}$        & -0.809   & -       \\
                          &         &       & $~t_{s^*s^*}$      & -4.241   & -       \\        
    \noalign{\smallskip}\hline
\end{tabular}
\end{table}
As we already mentioned, the surface reconstruction involves shifts of
surface atoms from their bulk position and leads to the formation of a
new Si-Si bond, absent in bulk. 
However, all Si atoms remain tetrahedrally coordinated and all bond
lengths are remarkably close to the bulk nearest-neighbor separation.
This is why for all Si atoms we utilize the TB parametrization
of Klimeck {\em et al.}~\cite{klimeck-IEEE2007} obtained by fitting the TB model to reproduce the experimentally observed bulk Si band structure.
The TB treatment of the H atoms passivating the top and the bottom Si
surfaces is performed by choosing the Si-H hopping
parameters identical to the Si-Si values, but adjusting the onsite
energy of the $s$ orbitals on each H atom to obtain a good fit to the
DFT data from Fig.~\ref{fig_bands_passsurface}(a). 
Values of all TB parameters are given in
Table~\ref{table_bulk_tb_parameters}.
We note that we have repeated the calculation of the band structure
accounting for the departures from bulk bond lengths and directions
by scaling all TB matrix elements with strain
corrections~\cite{klimeck-IEEE2007}.
The resulting band structure was very similar to that 
shown in Fig.~\ref{fig_bands_passsurface}(b), and so we chose to
utilize the bulk (unstrained) TB parametrization throughout this work.
As can be seen from Fig. \ref{fig_bands_passsurface}(b) this 
set of TB parameters captures all features of the band structure predicted by DFT. Our TB 
parametrization predicts larger electronic gaps for all symmetry points 
(see Table \ref{table_gaps_passivated}).
This is because the TB parameters were fitted to reproduce the
experimental  bulk Si bandstructure rather than that computed by DFT.

\section{ Dangling bond wires on a H:S\lowercase{i}-(100)-(2$\times$1) surface}
\label{subsec_dbwires}
\begin{figure}[b]
\begin{center}
\includegraphics[scale=0.9,angle=0]{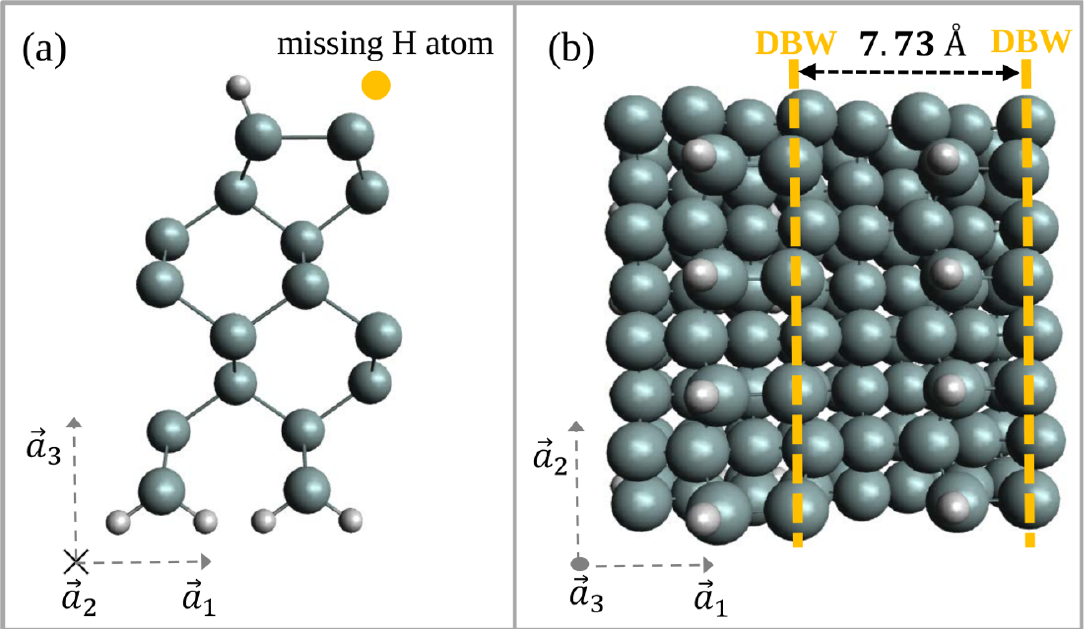}
\caption{ \label{fig_scell_dbw} (color online) (a) Side view of a supercell with a missing H
atom generating an array of infinite dangling bond wires (DBWs). (b) Top view of a larger 
supercell illustrating the formation of DBWs on parallel dimer rows separated by the lattice constant 
$a_1=7.73$ 
\AA.}
\end{center}
\end{figure}

If we remove a top H atom in the supercell shown in Fig. \ref{fig_scell_dbw}(a) and apply periodic 
boundary conditions, the resulting slab will have  an array of 
DB wires (DBWs) along the row of Si dimers separated by a single wire of 
saturated bonds and H atoms as shown in Fig. \ref{fig_scell_dbw}(b).  To validate our silicon computational box with reconstructed and passivated silicon surface , we compare here our DFT and TB results with DFT 
calculations reported in Ref. [\onlinecite{bohlou-guo-nanoletters-2016}], where the same configuration was studied. 
Before this comparison, we note that DBWs suffer both Peierls distortion and dimerization, and/or antiferromagnetic ordering, as discussed by Lee {\em et al.}   \cite{lee-zhang-prb2009,cho-kleinman-prb2002}. These dimerization effects are specific to the one-dimensional nature of the nanowire and, for the time being,  are beyond the scope of this work.
%
%

%
%
\begin{figure}[b]
\begin{center}
\includegraphics[scale=1.2,angle=0]{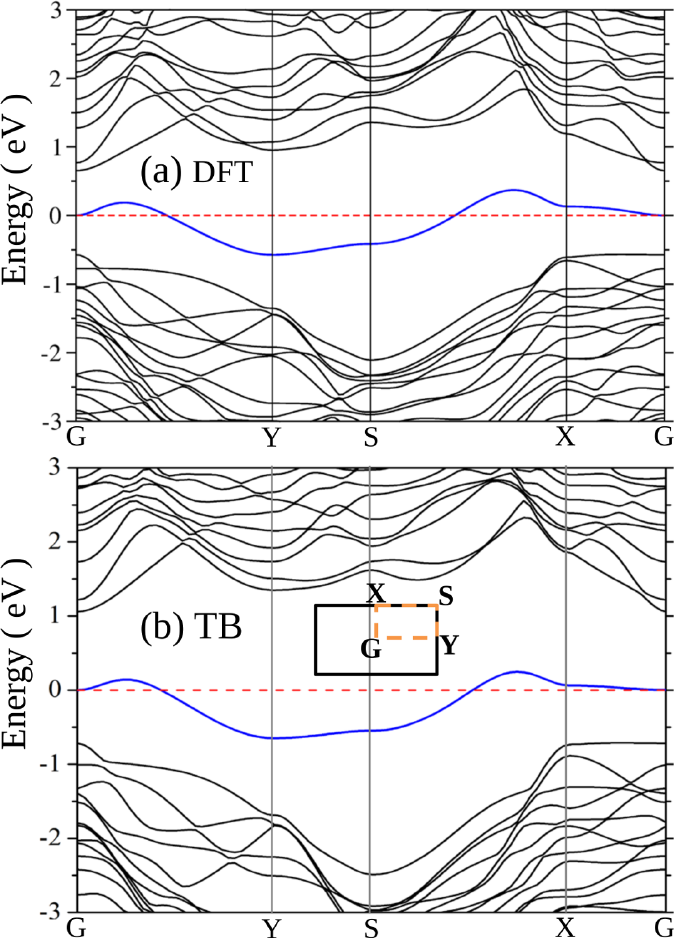}
\caption{ \label{fig_bands_dbw} (color online) (a) Energy band (in blue) in the Si bandgap 
associated with DBWs calculated with DFT (a) and with TB methods (b). The Fermi level 
is indicated with a dashed red line.}
\end{center}
\end{figure}

In Fig. \ref{fig_bands_dbw}(a) we show the results of DFT calculations of energy bands for the Si slab 
defined in Fig. \ref{fig_scell_dbw}. This supercell with a missing H atom has now an odd number of 
electrons $N_e=69$, hence we performed spin-polarized DFT calculations to obtain the band structure. 
We did not carry out the geometry optimization of the model slab. 
Such optimizations, performed by Watanabe {\em et al.} \cite{watanabe-wada-prb-1996}, revealed that 
the vertical position of the 
unsaturated Si atoms is lowered by only $0.1$ \AA.
We used the Fermi-Dirac smearing to define the occupations of spin KS orbitals at each k-point 
during self-consistent calculations. Converged results show no spin polarization, with the highest
occupied orbital equally populated by one half of the unpaired electron. 
This translates into identical band structures for the two spin components.

%
\begin{table}[h]
\caption{Energy gaps between the top of the valence band and the midgap band of DB states 
($E_\mathrm{gap}^\mathrm{DB}$) at points G, Y, S and X as plotted in Fig. \ref{fig_bands_dbw} 
calculated with DFT and TB. All values are given in eV.}
\label{table_gaps_dbw}
\begin{tabular}{ll|l}
\hline\noalign{\smallskip} 
  & $~~~~~$ DFT & $~~~~~~$ TB \\
\noalign{\smallskip}\hline\noalign{\smallskip}
$E_\mathrm{gap}^\mathrm{DB}(\mathrm{G})$ & $~~~~$ 0.570 $~~~~$   & $~~~~$ 0.716 $~~~~$ \\
                             &                       &                               \\
$E_\mathrm{gap}^\mathrm{DB}(\mathrm{Y})$ & $~~~~$ 0.778 $~~~~$   & $~~~~$ 1.036 $~~~~$ \\
                             &                       &                               \\
$E_\mathrm{gap}^\mathrm{DB}(\mathrm{S})$ & $~~~~$ 1.695 $~~~~$   & $~~~~$ 1.943 $~~~~$ \\
                             &                       &                               \\
$E_\mathrm{gap}^\mathrm{DB}(\mathrm{X})$ & $~~~~$ 0.738 $~~~~$   & $~~~~$ 0.795 $~~~~$ \\
\noalign{\smallskip}\hline
\end{tabular}
\end{table}

In Fig. \ref{fig_bands_dbw}(b) we plot the energy bands of the same Si slab obtained by solving the 
TB Hamiltonian, Eq. (\ref{eq_tb_hamilt}),  with periodic boundary conditions. In our 
TB calculations, the removal of the Hydrogen atom is accounted for by
 i) setting the hopping parameters between the unsaturated Si atom and removed H atoms to zero, and 
ii) shifting the onsite energy of the $s$ orbital centered on this H atoms up in energy to avoid indirect coupling with other 
atoms. Furthermore, we adjusted the onsite energies of $s$ and $p$ atomic 
orbitals of depassivated Si atoms to the values $\varepsilon_s=-1.7$ eV and $\varepsilon_p=2.04$ eV 
to improve the TB description of the midgap band of DB states with respect to DFT results. 

Both DFT and TB calculations show the emergence of a band of states in the Si bandgap,
associated with the DBWs. This band, marked in blue, comprises electronic 
states localized on Si atoms with DBs on the  surface. Note that this band 
shows a dispersion in all directions in reciprocal space due to coupling of DBs 
localized on different dimer rows. Furthermore, our DFT and TB calculations predict an 
energy-dispersion width for this band of $0.74$ eV which is in excellent 
agreement with previous calculations \cite{bohlou-guo-nanoletters-2016,doumergue-baratoff-prb-1999, watanabe-wada-prb-1996}. 
In Table \ref{table_gaps_dbw} we report the 
energy difference between the top of the valence band and the DB states at the principal symmetry points. 
Note  that TB calculations place these states at slightly higher energies as compared with 
analogous results calculated using DFT.

\section{Dangling bond quantum dot on the silicon surface}
\label{sec_db_qdots}
We now turn to the description of a single DBQD on a passivated and 
reconstructed Si surface. With a single DBQD we abandon the periodic boundary conditions in all 
directions
\begin{figure}[h]
\begin{center}
\includegraphics[scale=1,angle=0]{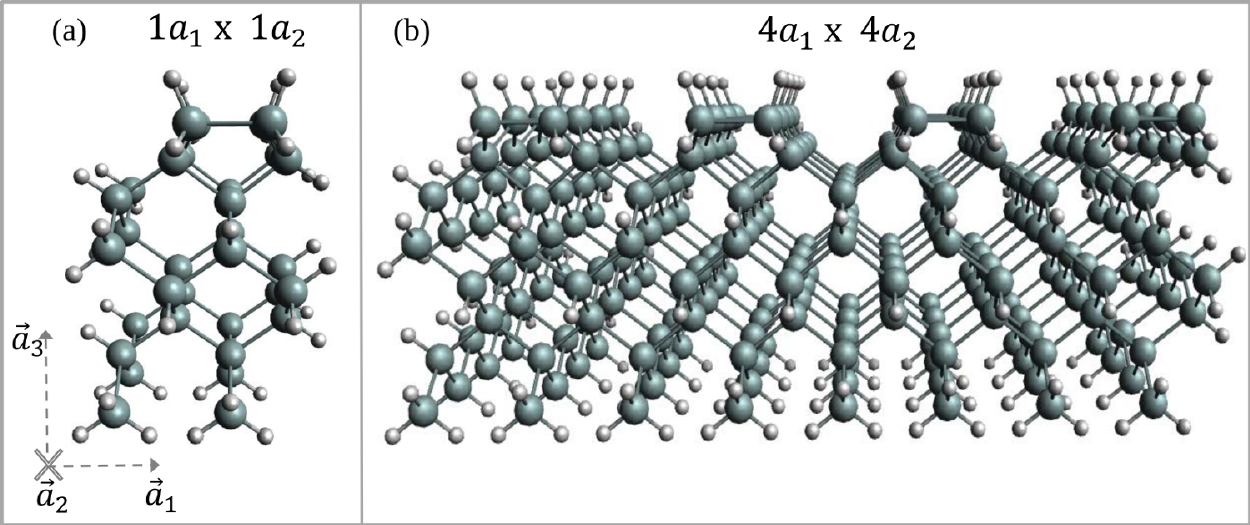}
\caption{ \label{fig_si_cbox} (color online) Silicon computational box (SiCB) with top 
reconstructed surface consisting of (a) $N_\mathrm{Si}=32$, $N_\mathrm{H}=40$, $N_e=168$ and (b) 
$N_\mathrm{Si}=256$, $N_\mathrm{H}=184$, $N_e=1208$. The edge and bottom surfaces are passivated 
with H atoms without reconstruction. SiCBs are labeled based on the number of unit cells ($n$) in 
the directions of ${\vec a}_1$ and ${\vec a}_2$ ($n a_1$ $\times$ $n a_2$).}
\end{center}
\end{figure}
%
and develop  a Si computational box (SiCB) properly passivated 
with H atoms on all sides, but with surface reconstruction only on the top side. 
Fig. \ref{fig_si_cbox} shows two computational boxes,  labeled by 
the number of unit cells ($n$) in the directions of ${\vec a}_1$ and ${\vec a}_2$ ($n a_1$ $\times$ 
$n a_2$). 
Fig. \ref{fig_si_cbox}(a) shows the smallest possible slab of 
$N_\mathrm{Si}=32$ Si atoms passivated with H. 
As already mentioned, the top layer is the reconstructed and 
passivated surface, while the unreconstructed bottom surface and edges are fully saturated with H 
atoms with the goal of simulating a seamless transition to the bulk Si. 
For $N_\mathrm{Si}=32$ Si atoms this requires $N_\mathrm{H}=40$ H atoms, with a total number 
of electrons $N_e=168$. In Fig. \ref{fig_si_cbox}(b), a much larger SiCB, with 
$N_\mathrm{Si}=256$, $N_\mathrm{H}=184$, and $N_e=1208$, is shown.

\begin{figure}[h]
\begin{center}
\includegraphics[scale=1,angle=0]{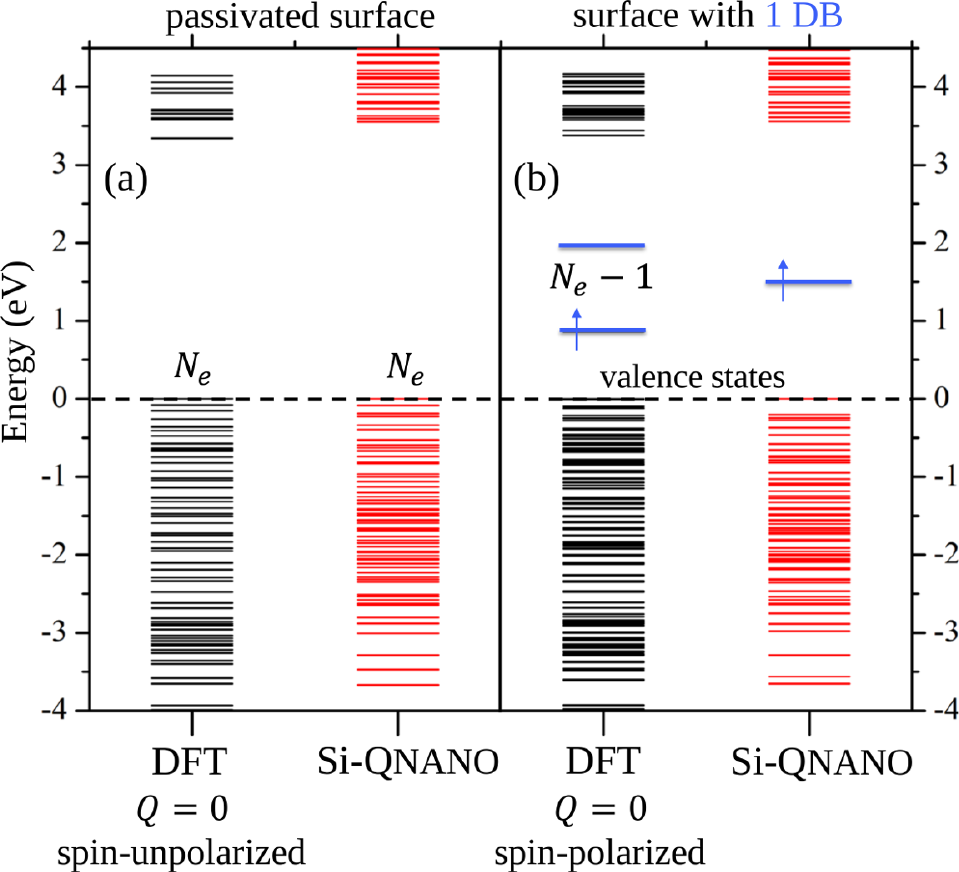}
\caption{ \label{fig_si_1x1_spectra} (color online) (a) Energy levels of the Si computational 
box (SiCB) shown in Fig. \ref{fig_si_cbox}(a) with the H-passivated top reconstructed surface.
$Q$ denotes the net charge of the SiCB. (b) The effect of removal of a H atom from a 
top surface and a formation of a single dangling bond state with energy in the Si bandgap.}
\end{center}
\end{figure}

Since these SiCBs are finite clusters, we carry out {\it ab-initio} DFT 
calculations using the {\sc Octopus} code \cite{ref_octopus_2015} to solve the KS equations 
in a real-space representation. The real-space simulation domain was defined by using spheres 
centered at the atomic positions with radius $5$~\AA{} and a uniform spacing of 
$0.19$~\AA{} between each grid point of the generated mesh. We used atomic pseudopotentials 
to account for core electrons of Si atoms and the PBE-GGA approximation to the 
exchange-correlation potential. We followed these simulations by TB calculations with 
Si-{\sc Qnano} with the parameters optimized to reproduce DFT energy bands of a Si slab as discussed 
in the previous sections.

Fig. \ref{fig_si_1x1_spectra} shows the results of DFT and TB 
calculations for the  $N_\mathrm{Si}=32$ Si slab. 
Fig. \ref{fig_si_1x1_spectra}(a) shows the energy levels of the slab fully passivated with 
H. Black bars show KS orbital energies calculated with {\sc Octopus} and red bars show 
the TB energy levels obtained with Si-{\sc Qnano}. We fill
the energy levels with spin 
up/down electrons up to the Fermi level which we take as reference energy level $E_F=0$. The 
structure is fully passivated, and we find the energy gap of $E_\mathrm{gap} \approx 3$ eV opening in 
the energy spectrum in both DFT and TB calculations. 
Fig. \ref{fig_si_1x1_spectra}(b) shows the 
effect of a removal of a H atom from the top surface, resulting in  
the formation of the DB. 
This implies a removal of one electron, and there are now 
$N_e=168 - 1$ electrons, with one electron on a Si DB. In addition, we 
characterize the cluster by the net charge $Q$. Because we removed 
both a proton and an electron, the net charge remains $Q=0$ even though there is a DB
and an odd number of electrons.

\begin{figure}[h]
\begin{center}
\includegraphics[scale=1.2,angle=0]{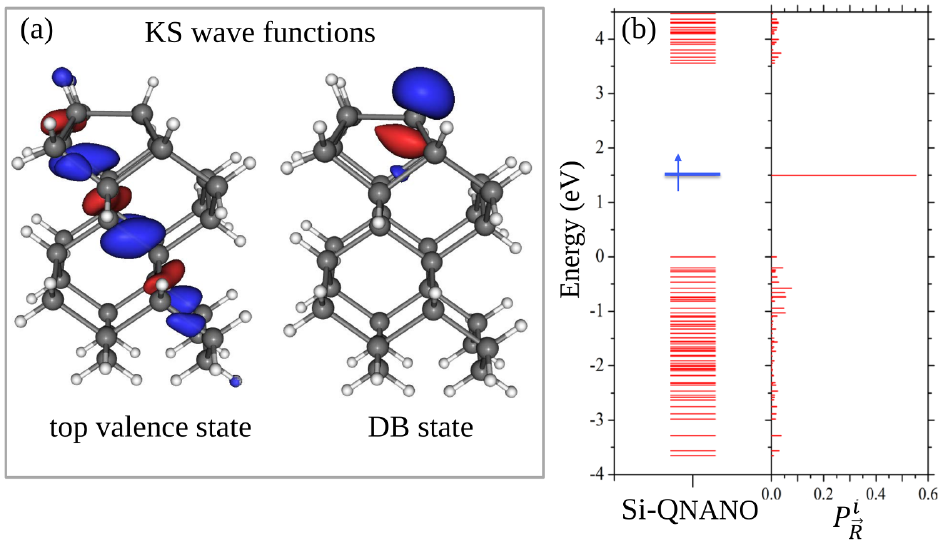}
\caption{ \label{fig_db_pd} (a) Isosurfaces of Kohn-Sham  wave functions of the top valence 
state and the DB state with energy in the gap. (b) Energy levels of the SiCB $1a_1\times1a_2$ shown in Fig. \ref{fig_si_cbox}(a) and probability 
density $P_{\vec{R}}^i$ of finding the electron localized around the depassivated Si atom with 
position vector $\vec{R}$ calculated for each TB state $\vert \phi_i \rangle$.}
\end{center}
\end{figure}

The energy spectra for a surface with one DB show an
emergence of an energy level, marked in blue, in the energy gap. As we have an odd number of 
electrons, in spin-polarized DFT the energy levels for spin up and spin down electrons are different.
Thus, we have two localized levels that appear at $0.9$ eV and $1.9$ eV above the top valence 
state, respectively. In red we show the energy spectrum for the same Si cluster with a DB 
state obtained with Si-{\sc Qnano}. We see that, just like in DFT calculations, there is a bound 
state, marked in blue, in the energy gap located at $1.5$ eV above the top valence state.

The localized nature of the DB state is shown in Fig. \ref{fig_db_pd}. In its left 
panel we visualize the KS wave functions of the top valence and DB states for the 
SiCB shown in Fig. \ref{fig_si_cbox}(a). We find that while the valence state 
delocalizes over several Si atoms of the cluster, the wave function associated with the DB
state is clearly localized around the Si atom with the DB. Furthermore, we also observe the $p_z$ atomic-like character of the DB KS orbital evident from the plotted isosurface . 
A similar 
scenario is confirmed by TB calculations. In Fig. \ref{fig_db_pd}(b) we plot the 
probability density $P_{\vec{R}}^{i}=\sum_\alpha \vert C_{\vec{R}\alpha}^i \vert^2$ of 
finding the electron localized around the Si atom with position vector $\vec{R}$ and with a DB, 
calculated for each eigenstate $\vert \phi_i \rangle$. We 
see a distinct peak of $P_{\vec{R}}^i$ associated with the state in the energy gap indicating a 
strong electron localization on the Si atom with the DB.

\begin{figure}[t]
\begin{center}
\includegraphics[scale=1.5,angle=0]{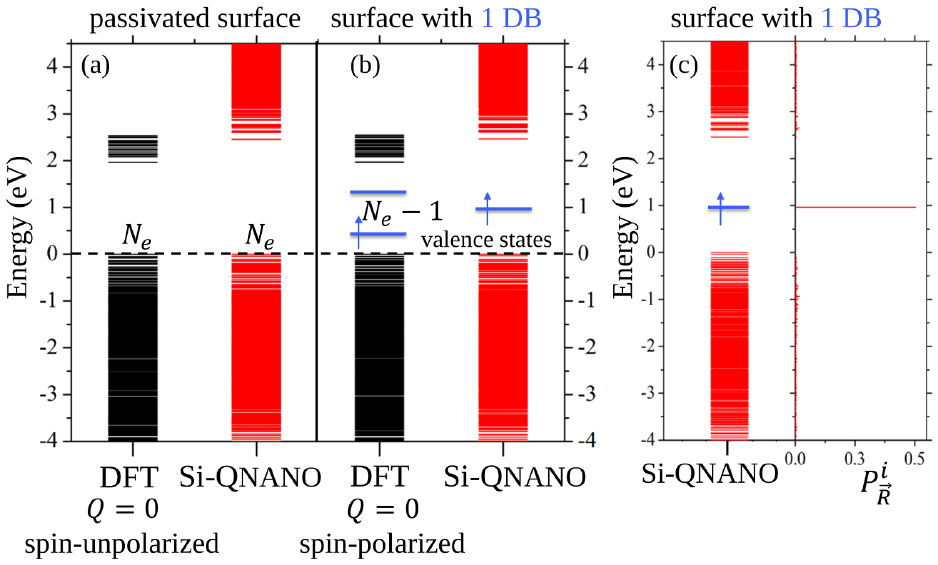}
\caption{ \label{fig_si_4x4_spectra} (a) Energy levels of the  SiCB $4a_1\times4a_2$ shown in Fig. \ref{fig_si_cbox}(b). $Q$ denotes the net charge of the SiCB. (b) 
The effect of removal of a H atom from the top surface and a formation of a single DB state with energy in the Si bandgap. (c) probability density $P_{\vec{R}}^i$ of finding the 
electron localized around the depassivated Si atom at position  $\vec{R}$, calculated for 
each TB state $\vert \phi_i \rangle$.}
\end{center}
\end{figure}

In Fig. \ref{fig_si_4x4_spectra} we show the results obtained for a much larger computational box with
$N_\mathrm{Si}=256$ and $N_\mathrm{H}=184$. We obtain qualitatively the same result. 
The energy gap in the spectrum appears, with the value $E_\mathrm{Gap} \approx ~2$ eV, which is smaller 
than the energy gap for the smaller cluster as shown in Fig. \ref{fig_si_1x1_spectra}. The removal of a 
H atom from the top surface results in the appearance of a DB state in both DFT and TB
spectra.
Its energy is respectively $0.4$ eV and $0.9$ eV above the top valence state.

\begin{figure}[h]
\begin{center}
\includegraphics[scale=1,angle=0]{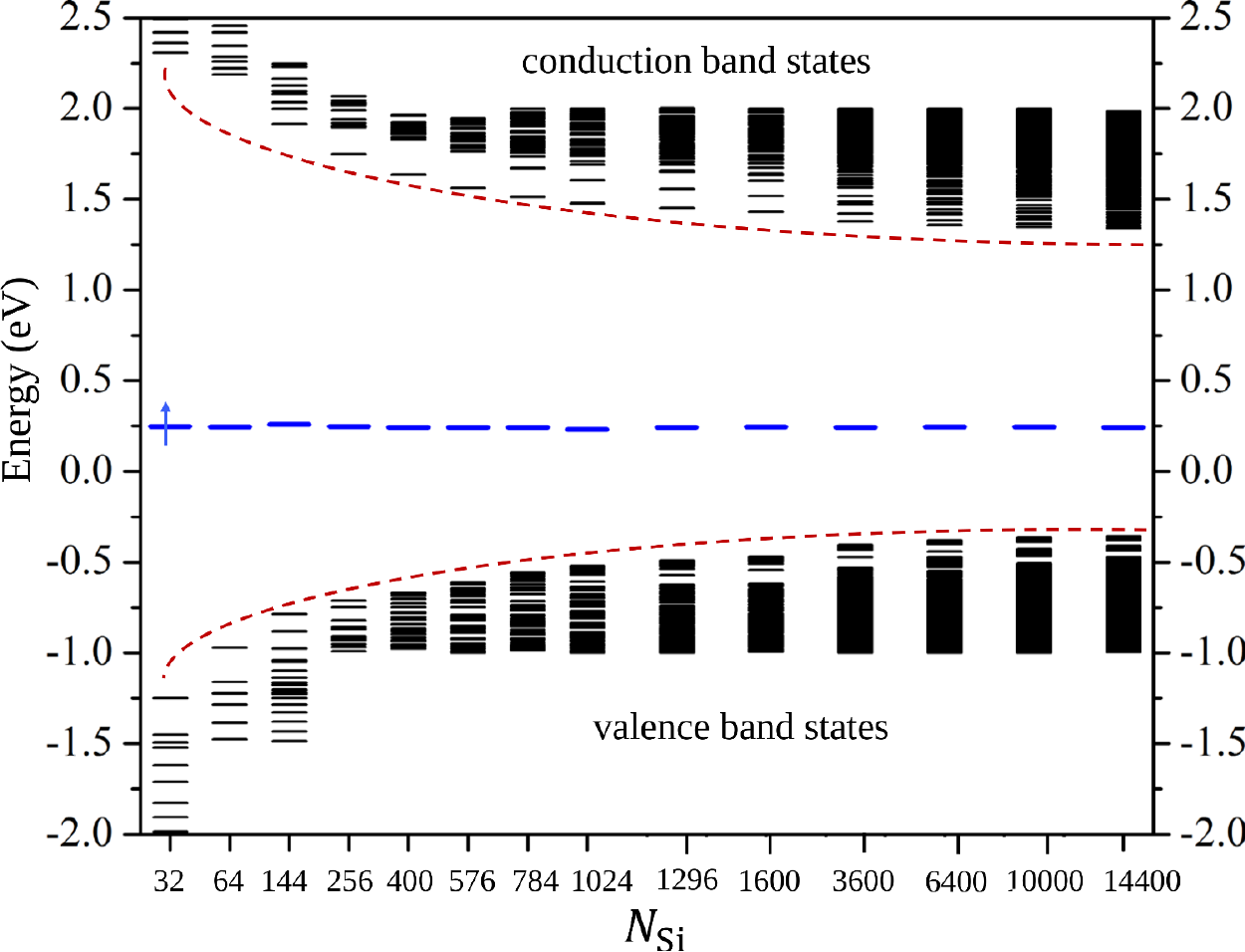}
\caption{ \label{fig_tb_spectra_sicbs} (a) Energy levels of  SiCBs of increasing size with a single DB in the top 
reconstructed surface calculated with Si-{\sc Qnano}. The DB state is marked in blue, the arrow 
indicates single-electron occupation of this state. Dashed red lines illustrate the 
evolution of the top of the valence band and the bottom of the conduction band with the size of 
the finite slab.}
\end{center}
\end{figure}

In Fig. \ref{fig_tb_spectra_sicbs} we plot the TB energy spectra calculated with Si-{\sc 
Qnano} of Si computational boxes of increasing size with a single DB at the top surface.
The simulated slabs consist of a total number of Si atoms $N_\mathrm{Si}$ ranging from 32 (the smallest 
possible cluster) to 14400, a SiCB with dimension $30a_1 \times 30 a_2$. 
We used the Lanczos method, as implemented in Si-{\sc Qnano}, to calculate the energy levels 
of the finite slabs close to the top of the valence band and to the bottom of conduction 
band. In all cases a gap opens with the appearance of a DB state with the energy in the gap. 
Dashed red line in Fig. \ref{fig_tb_spectra_sicbs} shows the renormalization of the valence and 
conduction band edges as the size of the system increases. The energy gap decreases, 
as expected for a less confined system, and its value for the largest SiCB, $E_\mathrm{Gap}=1.6$ eV, 
approaches the TB energy gap of an infinite slab as shown in Fig. \ref{fig_bands_dbw}. We 
also observe that the energy of the DB state converges very rapidly with the size of the Si slab 
due to the localized character of this state. For the largest SiCB with $N_\mathrm{Si}=14400$ atoms, TB 
calculations with Si-{\sc Qnano} place the DB state energy level $0.6$ eV above the top of 
the valence band.

\section{Charging energy of the dangling bond quantum dot}
\label{sec_charged_db_qdots}
STM experiments show that the dangling bonds simulated in the previous 
section are on average neutral, that is, the DB quantum dot is occupied by a single 
electron. However, for n-doped Si samples, additional electrons in the conduction band can be loaded into the DBQD in a controlled way using gates \cite{wolkow-bruno-2014-nanocomputing,taucer-wolkow-prl-2014}. 
We approximate the two-electron DBQD with a single configuration where two electrons with opposite 
spins are populating the DB orbital, with the net DBQD charge $Q=-1$. The DBQD
charging energy  $U_\mathrm{DB}$ is given by the two-body Coulomb matrix element,
\begin{eqnarray}
&& \kern-10pt U_\mathrm{DB} = \int d{\bf x}_1 d{\bf x}_2 ~ \phi_u^*({\bf x}_1) 
\phi_d^*({\bf x}_2) 
\frac{e^2}{ \epsilon({\bf x}_1, {\bf x}_2) \vert {\bf x}_1 - {\bf x}_2 \vert} \phi_d({\bf x}_2) 
\phi_u({\bf x}_1), 
\label{coulomb_me_1}
\end{eqnarray}
where $\phi_u$ and $\phi_d$ are spin-up and spin-down DB spin-orbitals, each occupied by 
one electron at position  ${\bf x} \equiv (\vec{r})$, and 
$\epsilon({\bf x}_1, {\bf x}_2)$ is the position-dependent dielectric function. In order to 
evaluate this Coulomb matrix element we express the TB orbitals in the LCAO basis,
\begin{equation}
 \vert \phi_i \rangle = \sum_{\vec{R},\alpha} C_{\vec{R}\alpha}^{i} ~ \vert 
\vec{R}\alpha\rangle \vert 
\chi_{m_\alpha}\rangle,
 \label{atomic_basis}
\end{equation}
with $\chi$ denoting the spin wave function. 
From Eq. (\ref{atomic_basis}) it follows that 
\begin{widetext}
\begin{eqnarray}
U_\mathrm{DB} = \sum_{\vec{R}_1, \alpha_1} && \sum_{\vec{R}_2, \alpha_2} \sum_{\vec{R}_3, \alpha_3} 
\sum_{\vec{R}_4, \alpha_4} 
C_{\vec{R}_1 \alpha_1}^{u^*}~C_{\vec{R}_2 \alpha_2}^{d^*}~C_{\vec{R}_3 \alpha_3}^{d}~C_{\vec{R}_4, 
\alpha_4}^{u} \nonumber \\
&& \times ~ \delta_{m_{\alpha_1},m_{\alpha_4}}\delta_{m_{\alpha_2},m_{\alpha_3}} 
\langle \vec{R}_1\alpha_1,\vec{R}_2\alpha_2
\left|\frac{e^2}{\epsilon(\vec{r}_1, \vec{r}_2)\left|\vec{r}_1-\vec{r}_2\right|}\right|
\vec{R}_3\alpha_3,\vec{R}_4\alpha_4 \rangle.
\label{coulomb_me_2} \\ \nonumber
\end{eqnarray}
\end{widetext}
Equation (\ref{coulomb_me_2}) contains integrals up to four-center; in what follows We restrict the terms
to one-center and two-center only. The integrals involving 
atomic orbitals centered on the same atom and on nearest-neighbor (NN) atoms are 
calculated numerically using Slater-type orbitals to approximate the radial part of the wave 
functions. The two-center integrals involving non-NN atoms 
are treated as long-range 
Coulomb interactions between distributions of two charges as explained in Ref. \cite{zielinski-hawrylak-prb-2010}.

As we have shown in Figs. \ref{fig_db_pd}(b) and  \ref{fig_si_4x4_spectra}(c), for the DB state 
the probability density of finding the electron localized around the depassivated Si atom peaks at
the large value of $0.6$. This implies that the onsite term in Eq. (\ref{coulomb_me_2}) will 
give most of the contribution to the charging energy $U_\mathrm{DB}$ of the DB quantum 
dot. Indeed, we found that the DB charging energy shows practically no dependence on the size 
of the SiCB. On the other hand, the position-dependent 
dielectric function accounting for screening effects due to the valence electrons and the Si-vacuum 
interface has been taken in an approximate form \cite{zielinski-hawrylak-prb-2010}.  More specifically, the on-site 
(dominant) contribution to the Coulomb interaction between the two electrons in the DB state is 
screened by the effective dielectric constant $\epsilon_\mathrm{eff}=(\epsilon_\mathrm{Si}+1)/2$ 
where $\varepsilon_\mathrm{Si}=11.9$ is the Si static dielectric constant
\cite{madelung-otfried-semiconductors}.
The smaller contributions involving atomic 
orbitals centered on neighboring atoms and non-NN atoms are screened by the dielectric constants 
$\epsilon_\mathrm{Si}/2$ and $\epsilon_\mathrm{Si}$, respectively.   

\begin{figure}[h]
\begin{center}
\includegraphics[scale=1.5,angle=0]{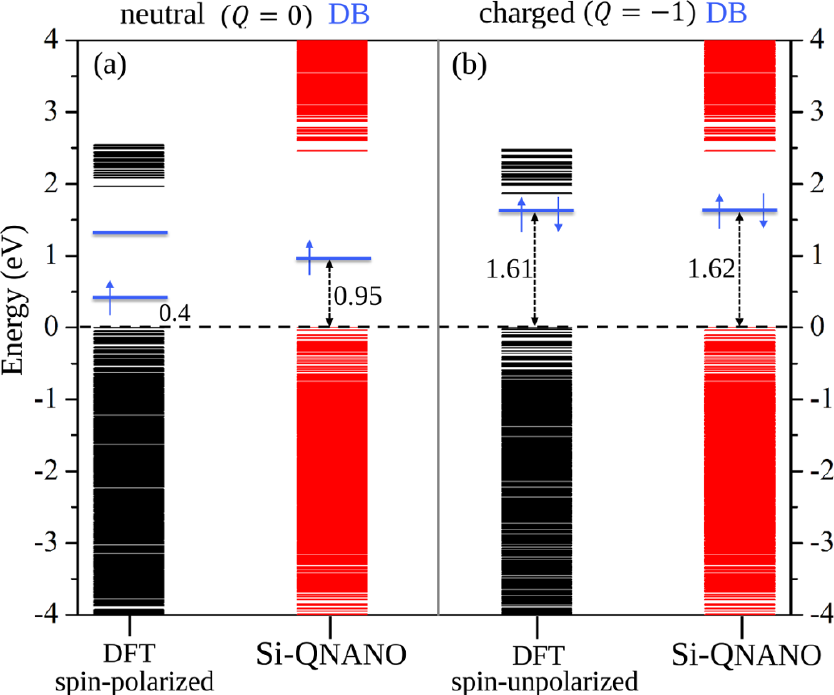}
\caption{\label{fig_charging_energy} Energy of the (a) neutral and (b) charged DB state for a 
SiCB $4a_1 \times 4a_2$ shown in Fig. \ref{fig_si_cbox}(b) with a depassivated Si atom at the 
top surface calculated with DFT and Si-{\sc Qnano}.}
\end{center}
\end{figure}

Fig. \ref{fig_charging_energy} shows the energy spectra of the SiCB shown in Fig. 
\ref{fig_si_cbox}(b) with a DB on the top surface calculated with DFT and the TB approach 
implemented by Si-{\sc Qnano}. The DB states shown in Fig. \ref{fig_charging_energy}(a) are 
occupied by one electron which corresponds to a neutral DBQD.  On the other 
hand, Fig. \ref{fig_charging_energy}(b) shows the energy spectra of SiCB with the net charge of $Q=-1$ 
where an extra electron with opposite spin populates the DB state. The DFT spectra are obtained by
solving the spin-unpolarized KS equations for the negatively charged finite Si box. 
Note that in this case the KS energy of the DB state is shifted up by $1.2$ eV with respect to the 
energy of the neutral DB state. This means that each electron occupying the DB state for the 
system with $Q=-1$ experiences a KS effective potential due to the delocalized valence electrons 
plus a two-body Coulomb repulsion due to the presence of a second electron in the DBQD.

Similar conclusions hold for electrons occupying the TB states calculated with Si-{\sc Qnano}. 
In this case, the energy of the DB state occupied by two non-interacting electrons will be 
shifted up in energy by the charging energy $U_\mathrm{DB}$ with respect to the singly-occupied DB state. In order to compare with DFT results, we used the TB wave function of the DB 
state to compute the charging energy using Eq. (\ref{coulomb_me_2}), for which we obtained the 
value of $0.67$ eV. This places the energy of the negatively charged DB state at $1.62$ eV above the 
top of the valence band which is in excellent agreement with DFT results as  shown in Fig. 
\ref{fig_charging_energy}(b). In other words, our TB approach predicts a $0.67$ eV upward shift for 
the energy level of the charged DB relative to that of the neutral DB. We note that this number is 
in good agreement with the value of $0.5$ eV reported by Livadaru {\it et al.} \cite{livadaru-wolkow-njp-2010}. Furthermore, the charged DB energy level predicted by 
Si-{\sc qnano} defines a bound state with energy $-0.85$ eV relative to the bottom of the 
conduction band which is similar to the value of $-0.95$ eV reported by Schofield {\em et al.} \cite{schofield-bowler-nature-2013} to model STM experimental results on charged DB qdots. 
The small deviations of our results relative to the data reported by other authors may be 
related to the fact that we did not re-optimize the atomic position of the depassivated Si atom due 
to the excess charge localization.

\section{CONCLUSION}
\label{sec_conclusions}
We presented Si-{\sc Qnano}, a new scalable computational platform to simulate atomic 
scale quantum devices in Si. The central result is the construction of a Si computational box with a reconstructed surface opening the way toward realistic simulation of atomic scale devices on Si surface.
We applied the Si-{\sc Qnano} computational box  to describe the dangling bond quantum 
dots on a hydrogen-passivated Si-(100)-(2x1) surface.
The dangling bond due to the removal of a H atom  was shown to result in an energy level localized in the Si bandgap, 
with wave function localized in the vicinity of the dangling bond silicon atom. 
The DBQD was shown to accommodate up to two electrons and the associated charging energy was predicted.  For small 
number of Si atoms, the Si-{\sc Qnano} results agreed very well with {\it ab-initio} 
calculations. However, Si-{\sc Qnano} allowed us to compute the electronic properties for Di
nanostructures involving tens of thousands of atoms. Future work will apply Si-{\sc Qnano} to dangling bond and 
dopant-based quantum circuits, gated quantum dots, and Si nanocrystals.
 
\begin{acknowledgments}
The authors thank R. Wolkow and K. Gordon of Quantum Silicon Inc., NSERC ENGAGE program and University of Ottawa Research Chair in Quantum Theory of Materials, 
Nanostructures and Devices for support.
\end{acknowledgments}

\bibliography{si_qnano_references}

\end{document}